\documentclass[11pt]{article}
\usepackage{amsmath,amsfonts}
\usepackage{fullpage}
\usepackage{lscape}

\def\.{\mbox{ \tiny{ $^\bullet$} }}

\def\eps{\epsilon}
\def\ko{k_0}

\def\les{\left[}
\def\ris{\right]}

\begin{document}

\begin{center}

{\large {\bf
On mediums with negative phase velocity:\\
a brief overview}}\\

\vskip 1 cm
\def\affil#1#2#3{\begin{itemize} \item[$^1$] #1 \item[$^2$] #2
\item[$^3$] #3
                  \end{itemize}}

Akhlesh Lakhtakia$^1$,\footnote{\noindent Email:
AXL4@psu.edu;
Telephone: +1 814 863 4319; Fax: +1 814 865 9974}
Martin W. McCall$^2$, Werner S. Weiglhofer$^3$,\\
Jaline Gerardin$^1$
and Jianwei Wang$^1$

\date{}

\affil
{CATMAS~---~Computational and Theoretical Materials
Sciences Group\\
Department of Engineering Science and Mechanics\\Pennsylvania
State University,
University Park, PA 16802--6812, USA}
{Department of
Physics, The Blackett Laboratory\\Imperial College of Science,
Technology and Medicine\\Prince Consort Road, London SW7 2BW, United
Kingdom}
{Department of Mathematics, University of Glasgow\\
Glasgow G12 8QW, United Kingdom}

\end{center}

\begin{abstract}
Several issues relating to oppositely directed
phase velocity and power flow  are reviewed. A necessary condition
for the occurrence of this phenomenon
in isotropic dielectric--magnetic
mediums
 is presented. Ramifications for aberration--free lenses,
homogenization approaches, and complex mediums are
discussed.\\

\noindent {\bf Keywords:} Left--handed materials,
Negative phase velocity, Negative
real permeability, Negative real permittivity
\end{abstract}

\section{Introduction}

As witnessed by the introduction of a session on the
so--called {\em left--handed materials\/} in this
conference, materials with negative phase velocity have
attracted much attention during the last two years.
Over three decades ago,
Veselago \cite{Ves}
suggested many unusual properties of  materials with
negative real relative permittivity and
negative real relative permeability at a certain frequency,
including
inverse refraction, negative radiation pressure, inverse Doppler
effect. However, his considerations were completely
speculative in view of the lack of a material, or even a
nonhomogeneous composite medium, with a relative permittivity
having a negative real part and a very small imaginary part. A
breakthrough was achieved by Smith {\em et al.\/} \cite{Schultz1},
who, developing some earlier ideas by Pendry {\em et al.\/}
\cite{Pendry1}--\cite{Pendry5}, presented evidence for a weakly
dissipative composite medium displaying negative values for the
real parts of its {\em effective\/} permittivity and {\em
effective\/} permeability. Their so--called {\em meta}--material
consists of various inclusions of conducting rings and wires
embedded within printed circuit boards.
Other types of nanostructural combinations with similar response
properties can also be devised \cite{dewar}.

Experimental results published last year by Shelby {\em et al.\/} \cite{SSS}
on the transmission of a $\sim 10$--GHz beam through a wedge
provided impetus
to the electromagnetics community
for a discussion  on the concept of a
negative index of refraction. Doubts have emerged
on the homogeneity and the isotropy of the composite materials used
by Shelby {\em et al.\/} as well as on the adequacy of
their measurement setup \cite{Lakh2}--\cite{GNV1}.
Despite those doubts,
the crucial flipping of the transmission pattern about the
normal to the exit surface, when a teflon wedge
was replaced by a wedge made of the ring--wire material,
appears to be unexplainable in any way other than
by resorting to the essence of Veselago's suggestion. In view of
the considerable literature accumulated during the
past few months and the explosive nature of the current
scientific scene \cite{Press}, we take this opportunity
to present our thoughts on a variety of related issues.

\section{What's in a name?}

The emergence of a clear terminology is often a difficult process
with regards to scientific findings relating to novel effects,
something that is also apparent in the present instance. The first
label for the candidate materials is {\em left--handed
materials\/} \cite{Ves}. But chiral materials are important
subjects of electromagnetics re\-search and the terms {\em
left--handedness\/} and {\em right--handedness\/} have been
applied to the molecular structure
   of such materials for well over a century \cite{SPIE1}.
The continued use of the term {\em left--handed materials\/}
(LHMs) for achiral materials
will thus confuse the crucial issues \cite{Schultz1,SSS,Smith,MS}.

The term {\em backward\/} (BW) medium  has been proffered
by Lindell and colleagues
\cite{Lindell}. This term presumes  the {\em a~priori\/} definitions of
forward and backward directions. Whatever be the merits
of this term for planewave propagation, it would founder
for problems involving nonplanar interfaces.

Ziolkowski and Heyman \cite{Ziol}
recently provided the most extensive theoretical
and numerical analysis of the negative index of refraction to
date. They  introduced the technical term {\em double
negative\/} (DNG)  medium  to indicate that the real parts of both
permittivity and permeability are negative. While sensible enough,
such nomenclature conceals the importance of dissipative effects.

After a careful study of the relevant
constitutive parameters, we have come to the conclusion
 that the term
{\em negative phase--velocity\/} (NPV) medium is unambiguous and
covers all possible situations that we could think of.
It also provides a contrast to the emerging
{\em negative group--velocity\/} (NGV) mediums, reports
on which are now emerging with regularity \cite{WKD}--\cite{KTM}.

\section{The condition for NPV}
Consider an isotropic dielectric--magnetic
medium with  relative permittivity
$\epsilon_r=\eps'_r+i\eps''_r$ and relative permeability $\mu_r
= \mu'_r+i\mu''_r$. Dissipation is reflected in the imaginary parts
$\epsilon''_r$ and $\mu''_r$, whilst causality dictates that
$\mu''_r
>0$ and $\epsilon''_r >0$, so that $\epsilon_r$ and $\mu_r$ lie in
the upper half of the complex plane.
The phase velocity is opposite to the
direction of power flow, whenever the inequality
\begin{equation}
\label{inequality} \left [ +{\left( {\eps'_r}^2+{\eps''_r}^2
\right )}^{1/2} - {\eps'_r} \right ] \left [ +{\left( {\mu'_r}^2
+{\mu''_r}^2   \right )}^{1/2}- {\mu'_r} \right ] >
\eps''_r\mu''_r~~
\end{equation}
holds. \cite{MLW02}
Clearly, the simultaneous satisfaction of
 both $\eps'_r <0$ and
$\mu'_r<0$ is a sufficient, but not necessary, requirement  for
the phase velocity to be negative. This result has been
illustrated by frequency--domain \cite{MLW02} as well as
time--domain \cite{WL02} calculations of planewave reflection at
the planar interface of free space and a NPV medium with
Lorentzian characteristics. 

A plane electromagnetic wave polarized
parallel to the $x$ axis,
and propagating along the $z$ axis in
a medium characterized by ${\epsilon}_r$ and $ {\mu}_r$, is described by
\begin{equation}
{\bf E}(z) = A\, \exp (i\ko n_{\pm} z) \, {\bf u}_x \,,
\end{equation}
where $n_{\pm} = \pm \sqrt {\epsilon_r \mu_r }$.
The choice of the sign of the refractive index is
mandated by the direction of power flow.
If the criterion
(\ref{inequality}) is satisfied, then $n_+$ (resp. $n_-$) 
applies for power flow along the $+z$ (resp. $-z$)
axis; accordingly,
${\rm Re}\les n_+\ris < 0$ (resp. ${\rm Re}\les n_-\ris > 0$), where
${\rm Re}\les\.\ris$ denotes the real part. Thus,
the phase velocity is oriented parallel to the $-z$ (resp. $+z$)
axis.  Some confusion in the literature \cite{PL1} emerges from the
claimed inadmissability of either $n_{+}$ or $n_{-}$ on grounds
other than the dictates of power flow.

\section{Perfect lenses}

Pendry \cite{PL1} presented  the possibility of fabricating a perfect
(i.e., aberration--free)
lens from a material with $\eps_r =\mu_r=-1$, which attracted
enormous attention from such luminaries as science
reporters attached to various newspapers. Attention
 came from researchers as well \cite{VWV,Ziol}, \cite{PL1a}--\cite{GNV2}.
In particular, Ziolkowski and Heyman \cite{Ziol} concluded
from extensive two--dimensional simulations that the condition
 $\eps_r =\mu_r=-1$ cannot be met by realistic meta--materials,
even in some narrow frequency range.

Aberrations due to dissipation inside
the desired NPV medium (with  $\eps_r =\mu_r=-1$)
would prove to be a stumbling
block in fabricating the desired perfect lenses \cite{Lakh0,GNV2}.
 However, the
possible use of active (i.e., non--passive) elements in
meta--materials may provide some relief from the glorious tyranny of the
principle of conservation of energy \cite{blah}. Chromatic aberrations due to
non--fulfilment of the required conditions outside some narrow
frequency range will also be important \cite{VWV,Lakh0}.

\section{Distributed Bragg reflectors}
The Bragg regime of a multilayer distributed Bragg reflector (DBR)
would undergo a blue--shift,
if a conventional positive phase--velocity (PPV)
constituent were to be replaced by its NPV counterpart \cite{GL02}.
An underlying cause may be the reversal of phase of the reflected
and the transmitted plane waves, when a PPV medium is replaced by
a NPV medium \cite{Lakh1}. Anyhow, multilayer DBRs with NPV
constituents could be useful in wavelength regimes that are
inaccessible with DBRs employing PPV constituents exclusively.

\section{Homogenization}
The meta--materials wherein the phase velocity can be
directed opposite to the power flow are composite materials
comprising inclusions of various kinds dispersed
in some host medium. At sufficiently low frequencies,
a homogeneous medium can be prescribed as {\em effectively\/}
equivalent to a particulate composite medium for
certain purposes \cite{SPIE2}--\c{W00}.

Incorporation of NPV mediums in such well--known homogenization
approaches as the ones named after Maxwell Garnett
and Bruggeman \cite{SPIE2} is mathematically trivial, and
we forecast many theoretical
publications thereon. Whether or not the theoretical predictions
of those approaches will be physically realized is
another matter. We do, however, note that surprising
results can emerge from the consideration
 of NPV mediums in homogenization
approaches. For instances \cite{Lakh2},
\begin{itemize}
\item [(i)] the
Bruggeman approach forecasts that a
certain mixture of a NPV medium
with its prosaic PPV counterpart, both impedance--matched,
can function as the medium that Pendry deems desirable for fabricating
perfect lenses \cite{PL1}; and
\item[(ii)] the Maxwell Garnett approach
predicts that composite mediums with zero permittivity and zero
permeability can be made as electrically small NPV
inclusions dispersed randomly but homogeneously
in a PPV host medium.
\end{itemize}

\section{Complex mediums}
Oppositely directed phase velocity
and power flow are distinct possibilities
in complex mediums, as discussed by Lindell
{\em et al.\/} \cite{Lindell}, and could yield
 interesting phenomenons. As an example, the
circular Bragg phenomenon will be reversed in ferrocholesteric
materials with negative real permittivities and
permeabilities \cite{Lakh3}. Even in isotropic chiral mediums,
negative $\eps'_r$ and $\mu'_r$ will lead to a reversed circular
dichroism \cite{Lakh4}.

To conclude, this brief overview of various emerging
 issues relating to NPV mediums
is expected to stimulate different lines of thought
among the participants of {\em Complex Mediums III\/}.
In view of claims and counter--claims launched respectively by Shelby
{\em et al.\/} \cite{SSS} and  the detractors of their
experimental results, the scientific situation is very presently volatile.
Though  the commonplace Lorentz model
does allow the possibility of isotropic, homogeneous,
dielectric--magnetic materials exhibiting negative phase velocity \cite{MLW02},
our brief overview definitely does not contain the last word on the
topic of artificial materials acting similarly.

\end{document}